\documentclass[12pt,a4paper]{article}

\usepackage[british]{babel}

\usepackage[a4paper,top=2cm,bottom=2cm,left=2.5cm,right=2.5cm,marginparwidth=1.75cm]{geometry}

\usepackage{amsmath, bm}
\usepackage{graphicx}
\usepackage[colorlinks=true, allcolors=blue]{hyperref}
\usepackage{hyperref}
\usepackage[title]{appendix}
\usepackage{mathrsfs}
\usepackage{amsfonts}
\usepackage{booktabs} 
\usepackage{caption}  
\usepackage{threeparttable} 
\usepackage{algorithm}
\usepackage{algorithmicx}
\usepackage{algpseudocode}
\usepackage{listings}
\usepackage{enumitem}
\usepackage{chngcntr}
\usepackage{booktabs}
\usepackage{lipsum}
\usepackage{subcaption}
\usepackage{authblk}
\usepackage[T1]{fontenc}    
\usepackage{lineno}
\usepackage{csquotes}       
\usepackage{diagbox}

\usepackage{setspace}
\usepackage{titlesec}
\titleformat{\section} 
  {\normalfont\Large\bfseries}{\thesection.}{1em}{}
  
\usepackage{float}   
\usepackage{caption} 


\title{Developing a Model-Consistent Reduced-Dimensionality training approach to quantify and reduce epistemic uncertainty in separated flows}

\author[1,*]{Minghan Chu}
\affil[1]{\small Mechanics and Materials Engineering Department, Queen's University, Kingston, K7L 3N6, ON, Canada}
\affil[*]{Corresponding author: \texttt{17mc93@queensu.ca}}

\date{}  

\begin{document}
\maketitle

\begin{abstract}
This proposed work introduces a data-assimilation-assisted approach to train neural networks, aimed at effectively reducing epistemic uncertainty in state estimates of separated flows. This method, referred to as model-consistent training, ensures that input features are derived directly from physics-based models, such as Reynolds Averaged Navier Stokes (RANS) turbulence models, to accurately represent the current state of the flow. Autoencoders have been selected for this task due to their capability to capture essential information from large datasets, making them particularly suitable for handling high-dimensional data with numerous discretization points in both spatial and temporal dimensions. This innovative approach integrates the ensemble Kalman method to enhance the training process, providing a robust framework for improving model accuracy and performance in turbulent flow predictions.

\end{abstract}






\section{Introduction}
Simulations of unsteady separated flows are routinely used in turbomachinery engineering applications. Reynolds Averaged Navier Stokes (RANS) turbulence models serve as the accepted model for these simulations. While the simplifications in RANS models make them computationally inexpensive, they also introduce significant epistemic uncertainty, referred to as structural uncertainty, into the Reynolds stresses in RANS formulations \cite{di2020effect, mishra2019uncertainty, gori2022confidence}. This predictive uncertainty leads to limitations in the engineering design and need to be estimated for reliable and optimal designs. Currently, the Eigenspace Perturbation Method (EPM) \cite{iaccarino2017eigenspace} is the only approach to predict these uncertainties. However, due to its purely physics-based nature and the uniform perturbations to the Reynolds stress tensor, this method often results in unrealistically large uncertainty bounds, leading to overly conservative designs.

Over the past few years, data-driven methods have been rapidly adopted in fluid mechanics, thanks to the availability of large high-fidelity datasets from direct numerical simulations and experiments \cite{brunton2020machine}. Data-driven methods have emerged as a promising approach used to develop closure models dedicated to RANS and large eddy simulation (LES) models \cite{duraisamy2021perspectives, chung2021data,chung2022interpretable, han2022vector, zhou2022frame}. Most of these studies adopt \textit{a priori} approach to augment a RANS model, where the neural network is trained offline without modifying the underlying source code of the RANS equations. The concept of \textit{priori} training minimizes the loss (discrepancy) between the output of a neural network and high-fidelity DNS using the back-propagation method \cite{tracey2013application, ling2016machine, weatheritt2016novel, wu2018physics, beetham2021sparse}. After the training process, the trained model is coupled with the RANS or LES solver for prediction. Although the training process is nonintrusive (i.e., without involving the RANS or LES solver and constrained directly by extracting physics from DNS), the \textit{priori} training is notorious for the inconsistency between training and prediction \cite{duraisamy2021perspectives, zhang2022ensemble}. This inconsistency may cause unexpected errors in prediction due to the issue of ill-conditioning \cite{wu2019reynolds}. According to \cite{thompson2016methodology,wu2019reynolds,duraisamy2021perspectives}, even replacing RANS-based fields with exact turbulence fields from DNS did not result in satisfactory predictive outcomes. Additionally, the \textit{priori} methods require a full field of DNS data, which is the key limitation due to the unavailability of DNS data for real-world applications. 

To address this problem, the input features need to be generated from the RANS or LES predictions in a manner consistent with the loss function used for training. This is referred to as `model-consistent training' \cite{duraisamy2021perspectives}. For model-consistent training, the loss is minimized using field inversion approaches, such as 1) deterministic (exact) gradient-based methods (e.g., adjoint-based methods) \cite{wang2010quantification,duraisamy2015new,strofer2021ensemble} and 2) probabilistic Bayesian-based methods (e.g., ensemble Kalman methods) \cite{singh2016using, wu2019reynolds, yang2020improving}. Adjoint-based methods require accurate access to the full field of variables and gradients. However, notable drawbacks include \cite{duraisamy2021perspectives, zhang2022ensemble}: 1) significant complexity in implementation (e.g., deriving and coding adjoint equations can be mathematically and computationally challenging), and 2) time consumption due to the need for high accuracy and exact gradient computation. As an alternative to adjoint-based methods, a recent study by \cite{strofer2021ensemble} uses ensemble-based methods to approximate derivatives. The ensemble-based gradient method is nonintrusive and can treat the PDEs as a black box. This capability enables ensemble-based methods to address complex, non-differentiable, or noisy systems effectively. The drawback is ensemble based gradient method is less accurate than the exact adjoint-based method \cite{evensen2018analysis}. To improve accuracy the ensemble Kalman methods can be adapted to directly update the weights and biases of the neural network based on observed data.  The covariance matrix guides the update process, implicitly incorporating both first-order and second-order gradient information to accelerate convergence and improve the accuracy of the parameter updates. Neural network augmentation with the ensemble Kalman method is scarcely studied. Very recently, the study by \cite{zhang2022ensemble} proposed a method to train a supervised neural network using the ensemble Kalman method. This method utilizes ensemble predictions to estimate the covariance matrix, representing uncertainties in the state estimates.

Given the recent rapid development in computational power, high-dimensional data with a vast number of discretization points in both spatial and temporal directions are increasingly being utilized. Neural networks capable of handling these high-dimensional dynamical systems with large and complex datasets are attracting significant attention. Unsupervised autoencoders have emerged as a powerful tool for managing high-dimensional nonlinear dynamical systems due to their ability to capture essential information from large datasets. By compressing input data into a lower-dimensional latent space, autoencoders focus on the most critical patterns and features. However, studies on employing autoencoders in model-consistent training are still very limited.

Recent research has demonstrated the effectiveness of autoencoders in various applications. For instance, autoencoders have been used to distinguish the linear and nonlinear responses of airfoil pressure distribution to changes in the angle of attack \cite{saetta2022machine}. Furthermore, it has been shown that autoencoders can capture useful information on prediction confidence even in the absence of ground truth data \cite{saetta2024uncertainty}.

Autoencoders can be trained using the ensemble Kalman method to update their weights and biases, combining the strengths of autoencoders in dimensionality reduction and feature learning with the robust state estimation capabilities of the ensemble Kalman method. However, integrating autoencoders with the ensemble Kalman method in flow-physics problems is still in its infancy and requires further investigation.

Given this context, the primary goal of the proposed work is to develop a model-consistent reduced-dimensionality training method to effectively reduce epistemic uncertainty in the state estimates of separated flows. To achieve this goal, we propose three research objectives:

\begin{enumerate}
\itemsep0em
    \item Develop a model-consistent training framework that trains a reduced-dimensionality deep-learning model by incorporating observed data through a data assimilation method.
    \item Assess the performance of the trained neural network by evaluating the degree of uncertainty reduction and improvements in prediction accuracy.
    \item Inform management decision-making to lead to better design, overall performance, and safety of machinery.
\end{enumerate}

\section{Objective}
The overarching goal of this proposed work is to develop a rigorous ensemble-Kalman-autoencoder training framework. This framework aims to adjust the neural network parameters to better fit observed data, minimizing the discrepancies between predictions and actual observations, thereby improving overall model performance in quantifying and reducing epistemic uncertainty. In this proposed work, autoencoders will be chosen as the machine learning model to be trained through the ensemble Kalman method for the following three practical advantages:

\begin{enumerate}
    \item By incorporating observed data and estimating the distribution of weights and biases, ensemble Kalman method provides a robust way to train the autoencoder, potentially leading to better generalization and performance.
    \item Ensemble Kalman method can effectively handle uncertainties in the parameter estimates, making the training process more resilient to noise and variations in the data.
    \item The sample-based nature of ensemble Kalman method allows it to handle non-Gaussian and nonlinear dynamics, which are often present in complex data distributions encountered in autoencoder applications.
\end{enumerate}



The proposed work will develop an ensemble-Kalman-autoencoder training framework that leverages machine learning predictions, physics knowledge, and observations to yield accurate estimations of epistemic uncertainty. This framework will be tested for estimating quantities in turbomachinery separated flows, such as the separation location, drag coefficient, lift coefficient, and pressure coefficient. The three research objectives of the proposed work are as follows:

\begin{enumerate}[noitemsep]
     \item Development of an ensemble-Kalman-autoencoder framework. 
     \item Assess the performance in accurately capturing the behavior of turbomachinery separated flows, including separation location, drag coefficient, and pressure coefficient. Evaluate the framework's impact on uncertainty quantification capability and convergence speed.
     \item Enhance confidence in decision-making by improving the understanding of uncertainty quantification.
     
\end{enumerate}

\section{Approach}\label{sec2}

Problem Formulation: To inform decision-making for better design, I propose a framework that fuse autoencoders, RANS formulations, and observations. The method to quantify and reduce the structural uncertainties would eventually guide the design of optimal turbomachines (i.e., risk management, cost savings, and safety of machinery process) in routine engineering applications for management-relevant decisions. 


The Reynolds stress tensor can be decomposed into eigenvalues ($\lambda_{ij}$), eigenvectors ($v_{ij}$), and the amplitude ($k$). These components are not directly measurable and must be indirectly inferred from observed data. Therefore, they are referred to as \textit{decision-level states}.

The proposed ensemble-Kalman-autoencoder training framework includes several key components:

\begin{enumerate}
  \item The state vector $\mathbf{X}= (\alpha_{o}, \mathbf{k}, \mathbf{\lambda_{ij}}, \mathbf{v_{ij}})$, which represents the estimated observed state (i.e., $\alpha_{o}$) and hidden state variables (i.e.,  $\mathbf{k}, \mathbf{\lambda_{ij}}, \mathbf{v_{ij}}$).
  \item  A forward model that includes model uncertainties. Here, the forward model includes RANS equations and an autoencoder-based neural network with uncertainties arising from various sources, including data volume, data quality, learning rate, batch size, and the regularization coefficient. These uncertainties are propagated into the predicted Reynolds stress field (i.e.,$ \mathbf{k}, \mathbf{\lambda_{ij}}, \mathbf{v_{ij}}$).
  \item High-fidelity observations with artificially added noise/error covariance matrix $\epsilon$.
\end{enumerate}

The observation vector is denoted $\mathbf{Z_{d}}$ (where the subscript `d' stands for `data'), which can include state observations such as velocity or pressure drag. The ensemble-Kalman-autoencoder training can be formulated as a minimization problem of appropriate loss function $\mathbf{L}$. I will use an ensemble Kalman method for neural network training in the proposed work, considering the error covariance matrices as described in \cite{evensen2022data} for implementing the fusion algorithms. The fusion algorithms solve a state estimation problem, which can naturally be approached using Bayesian inference \cite{evensen2022data}. The observations enter the ensemble-Kalman-autoencoder problem via the likelihood $f(\mathbf{Z_d|X})$, where $f$ is the probability density function (pdf). Importantly, the likelihood $f(\mathbf{Z_d|X})$ reflects the pdf of observed errors $f(\mathbf{\epsilon})$, where $\mathbf{\epsilon}$ is the matrix containing observed errors, referred to as the error covariance matrix. Since I plan to use high-fidelity DNS observations, the values in the $\mathbf{\epsilon}$ matrix should be negligibly small. Specifying $\mathbf{\epsilon}$ is a key step in quantifying the effect of observed uncertainties in obtaining the estimated state $\mathbf{X}$.

\textbf{Technical Innovation:} In the proposed work, the observed data serves as constraints during training of the neural network.  This allows for a more general and accurate estimation of $\mathbf{X}$ via autoencoders  (Obj. 1, Task 1.1 and 1.2). Another key innovation is that the autoencoders will be optimized by adding a regularization term in the loss function, incorporating the observed data (Obj. 1, Task 1.3). This ultimately enables a probabilistic prediction of $\mathbf{X}$ in space and time, which is necessary to inform management decision-making, leading to better design, overall performance, and safety of the machinery (Obj. 2).


\subsection{Methodology}

\subsubsection*{Objective 1: Development of an ensemble-Kalman-autoencoder framework}

This objective will be carried out by completing the following three tasks.

Task 1.1: Learn a turbulence model represented by autoencoders

This proposed work will integrate an autoencoder with an ensemble Kalman framework, incorporating the observed data as a regularization term in the loss function (see Task 1.2). The ensemble Kalman framework assimilates the available observed data to infer the missing states of a system. In this work, an autoencoder will be trained to represent the Reynolds stress tensor (as modeled by a closure model), and the model parameters (weight vector $w$) of the autoencoder will be inferred through the ensemble Kalman approach. The procedure for model learning through the ensemble Kalman method consists of the following steps:”

\begin{figure} 
\centerline{\includegraphics[width=6in]{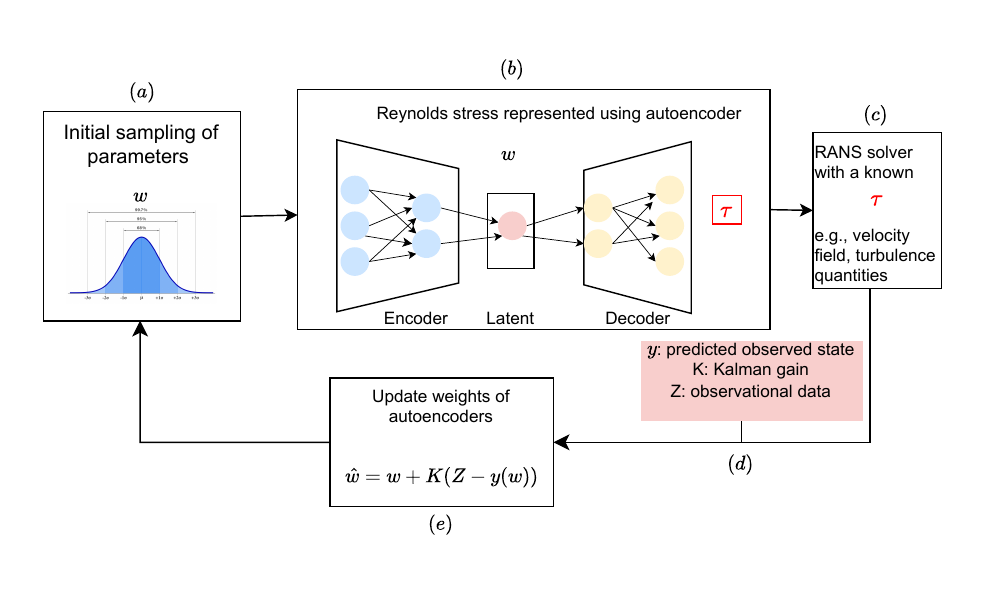}}
\caption{Schematic of autoencoder learning through an ensemble Kalman framework: (a) Initial sampling of weights for the autoencoder; (b) Autoencoder representation of a turbulence model; (c) Solving the velocity field using a RANS solver with the calculated autoencoder-based Reynolds stress field; (d) Using a RANS solver to predict the observed state, along with the observed data and calculated Kalman gain, to update the weights of autoencoders. (e) Updating autoencoder weights by assimilating observed data (i.e., Kalman gain ($K$) is a factor that determines the relative weight given to the new observed data versus the predicted observed state.).}
\label{fig:autoencoder_EnKF.pdf}
\end{figure}

\begin{figure} 
\centerline{\includegraphics[width=6in]{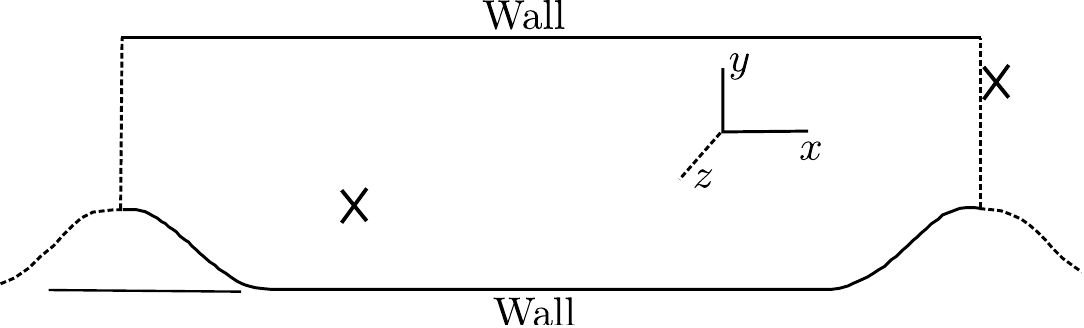}}
\caption{The 2D periodic hills computational domain. $\times$ symbols represent the locations from which observed data are extracted.}
\label{fig:periodichill.pdf}
\end{figure}

\begin{enumerate}
    \item The initial weights ($w$) for the autoencoder will be sampled from baseline distribution. (see Figure \ref{fig:autoencoder_EnKF.pdf} (a)). If we assume these parameters are independent and identically distributed (i.i.d), the initial weights can be defined using a Gaussian distribution $\mathcal{N}(w_0, \sigma^2)$ with mean $w_0$ and variance $\sigma^2$. 
    \item The autoencoder model will be used to represent the deviatoric part of Reynolds stress $\tau$. It consists of the hidden states of $k$, $\lambda_{ij}$, $v_{ij}$ (see Figure \ref{fig:autoencoder_EnKF.pdf} (b)). 
    \item The Reynolds stress term ($\tau$) appears in the momentum equation and can be solved using a RANS solver with a given velocity field (see Figure \ref{fig:autoencoder_EnKF.pdf} (c)). The initial velocity field will be obtained using a baseline turbulence model. 
    \item Using the RANS solver, the current predicted observed state is calculated. With the Kalman gain and the new observed data, the updated weights of the autoencoders can be determined (see Figure \ref{fig:autoencoder_EnKF.pdf} (d)).
    \item The Reynolds stress term is updated by updating the weights through the statistical analysis of comparing the predicted observed quantities (e.g., velocity field at specified locations or the drag from integrating the surface pressure) with observed data (e.g. sparse velocity observations, e.g., see Figure \ref{fig:periodichill.pdf}) using the ensemble Kalman method (see Figure \ref{fig:autoencoder_EnKF.pdf} (e)).  
\end{enumerate}


Task 1.2: Loss function in training the autoencoder uisng an ensemble Kalman framework

The goal of training is to minimize the loss $L = L(\boldsymbol{x}, g(f(\boldsymbol{x})))$, where $f$ and $g$ are nonlinear functions representing the encoder and decoder, respectively. The encoder maps the input $x$ to a latent representation $z$ (i.e., $z=f(x)$), and the decoder maps the latent representation back to the reconstructed input $\hat{x}$ (i.e., $\hat{x} = g(z)$ or $\hat{x} = g(f(x))$). The loss function can be a mean squared error (MSE) (i.e., $\|x-\hat{x}\|^2=(x-\hat{x})^2$) or even a mean absolute error (MAE) (i.e., $\|x-\hat{x}\|=(x-\hat{x})$). Therefore, reconstruction loss can be defined as:
 
 \begin{equation}
     L(x, g(f(x)))=\|x-\hat{x}\|^2=\|x-g(f(x))\|^2 \\
 \end{equation}

 or

 \begin{equation}
     L(x, g(f(x)))=\|x-\hat{x}\|=\|x-g(f(x))\|
 \end{equation}

To incorporate observed data, I will introduce a regularization term based on the observed data. This term penalizes the difference between the predicted observed quantity $y$ and the actual observation $Z$. In this context, $y$ represents the observed quantity (e.g., velocity field or drag), and $Z$ represents the available observed data (e.g., DNS data). The observation loss can be defined as:

\begin{equation}
    \mathrm{Observation \ Loss} = \|y-Z\|^{2}
\end{equation}

The total loss function combines the reconstruction loss and the observation loss. The observation loss acts as a regularization term that incorporates the observed data:

\begin{equation}
    \mathrm{Total \ Loss} = L(x, g(f(x))) + \lambda \|y-Z\|^{2}
\end{equation}.

Here, $\lambda$ is a regularization parameter that controls the trade-off between the reconstruction loss and the observation loss.

Task 1.3: Creating observations 

The proposed ensemble-Kalman-autoencoder framework indirectly assimilates observations to infer the hidden Reynolds stress field, requiring only measurable flow quantities such as sparse observations of mean velocities or integral quantities like drag and lift (e.g., see Figure \ref{fig:periodichill.pdf}). This approach eliminates the need to measure the full-field Reynolds stresses. To date, the role of measurement noise has often been investigated by adding artificial noise (e.g., uniformly sampled noise) to noise-free data \cite{maulik2020probabilistic}. The error matrix $\epsilon$ for the high-fidelity DNS data is expected to be small. In this work, the observed DNS data are obtained from in-house DNS data generated by a former PhD student from Queen's University. The noise values in the error matrix $\epsilon$ are negligibly small and will be constructed by adding Gaussian random noise (i.e., $\mathcal{N}(Z, \sigma_{z})$ and $\epsilon = (0, \sigma_{z}^2)$).

\subsubsection*{Objective 2: Inform management decision-making to enhance design effectiveness}
This objective outlines how integrating autoencoders with an ensemble Kalman method can enhance our understanding of uncertainty quantification, improve design, and ensure safety. The ensemble-Kalman-autoencoder algorithm developed in Objective 1 will inform management decision-making. The proposed work aims to consider both prediction accuracy and the degree of uncertainty reduction for the quantities of interest in management decisions. By simultaneously quantify these two metrics, we can decide when the decision time point will be to implement management or mitigation measures, and also what effective measure(s) to choose. 

The level of accuracy is assessed by comparing the predicted mean with reference data. As shown in Figure \ref{fig:hills_prior_posterior} (a), the posterior distributions at four different locations closely approximate the truth compared to the prior distributions. In addition to the mean, the uncertainty range indicates the confidence level that designers should place in the predicted mean. For instance, Figure \ref{fig:hills_prior_posterior} (a) illustrates a trend where sample distributions shrink in size compared to those in the prior distributions, signaling an increased confidence in the posterior distributions.

With a clear understanding of uncertainties, managers can make more informed decisions based on posterior probability distributions rather than solely deterministic predictions. This approach aids in optimal resource allocation by identifying where to focus efforts and resources to effectively mitigate risks. For instance, in Figure \ref{fig:hills_prior_posterior} (a), the posterior distribution at the far-right location of the geometry deviates significantly from the truth with a high degree of uncertainty. Recognizing this pattern can prompt designers to prioritize model redesign for flow simulation in that specific area of the computational domain. Robust designs are achievable only when designers comprehend the range and nature of uncertainties, allowing them to create systems that perform reliably under diverse conditions, not just ideal or expected scenarios. Understanding uncertainties also facilitates setting appropriate safety margins, ensuring designs remain safe even when faced with unexpected conditions.

\section{Preliminary tests and future work}

\begin{figure}
    \centering
    \includegraphics[width=0.9\textwidth]{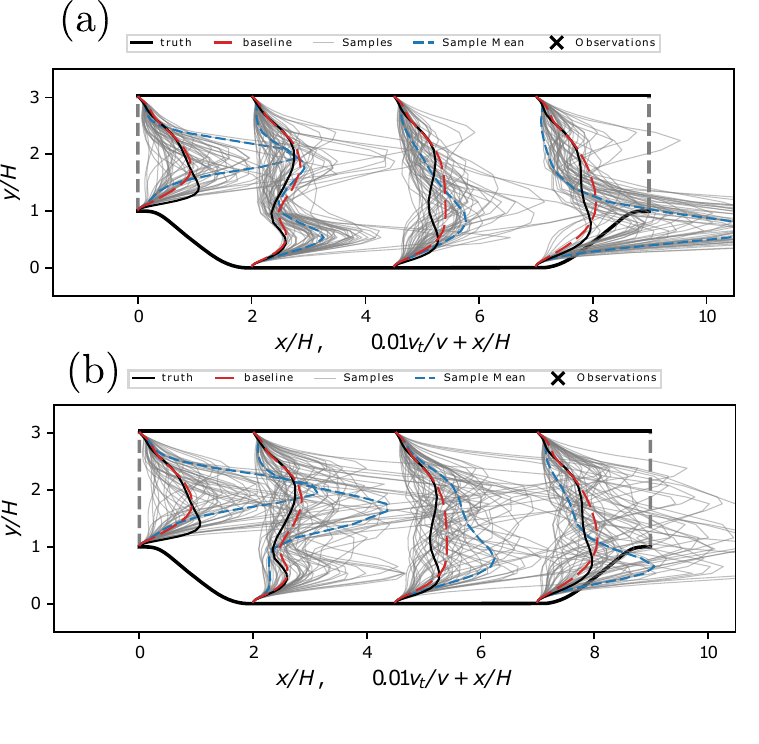}
    \caption{\textit{IEnKF}-based inference of Eddy viscosity (a) prior distributions (b) posterior distributions.
 }
    \label{fig:hills_prior_posterior}
\end{figure}

I have tested and validated the concept using the canonical case of flow over periodic hills. The periodic hill flow features a recirculation zone formed by forced separation, a strong mean flow curvature due to the domain geometry, and a shear layer that is not aligned with the overall flow direction. I used the open-source platform OpenFOAM \cite{openfoam} to solve a RANS-based $k-\omega$ model. First, the baseline RANS was solved using the built-in \texttt{simpleFOAM} solver for incompressible, steady-state turbulent flow simulations. As initial conditions, the generated baseline predictions were updated using the \textit{iterative Ensemble Kalman filter} (\textit{IEnKF}) method. The second forward RANS solver, \texttt{nutFoam}, was coupled with the \textit{IEnKF} framework to compute velocity with a given Reynolds stress field. Therefore, the \texttt{nutFoam} solver does not need to solve the equations for turbulence quantities, as the Reynolds stress field is reconstructed using Karhunen–Loeve (KL) expansions \cite{le2010spectral}. The combined OpenFOAM solvers and \textit{IEnKF} methods are implemented on the open-source DAFI platform by Xiao \textit{et al.} \cite{strofer2021dafi}. The data-assimilation framework assimilates the observations obtained from the marked locations, as shown in Figure \ref{fig:periodichill.pdf}, to infer both the observed velocity field and the hidden Reynolds stress field.
 

Figure \ref{fig:hills_prior_posterior} shows the predictions for the hidden state of eddy viscosity for flow over a periodic hill. For the sake of brevity, the case setup is omitted in this work; please refer to the paper by Xiao \textit{et al.} for details \cite{xiao2016quantifying}. From Figures (a) and (b), it is evident that the \texttt{IEnKF} method generally updates the posterior eddy viscosity profiles towards the true values, compared to its prior. However, an exception is observed within the recirculation region at the rightmost side of the geometry. 

Future work will develop a new ensemble-Kalman assisted approach that employs autoencoders to model the deviatoric part of the Reynolds stress tensor. Open-source software such as DAFI can serve as the foundation for implementations, for instance, by replacing the KL expansion with autoencoders. The deviatoric part of the Reynolds stress tensor can be expressed as $\tau_{ij}^{\mathrm{dev}} = 2k \sum_{n=1}^N c^{(n)} T_{ij}^{(n)}$, where $c^{(n)}$ are scalar coefficient functions that need to be determined, $T_{ij}^{(n)}$ are the tensor basis functions, and $N$ is the number of tensor basis functions used \cite{pope2001turbulent}.

\section{Conclusions}
The proposed work focuses on the development of a novel data-assimilation-assisted approach, integrating the ensemble Kalman method with autoencoders to address complex flow problems characterized by high degrees of turbulence in separated regions. By incorporating observed data as a regularization term in the loss function, this method aims to adjust neural network parameters to better fit actual observations, thereby minimizing discrepancies between predictions and real-world data. This approach leverages machine learning predictions, physics knowledge, and observations to yield accurate estimations of epistemic uncertainty.

The ensemble Kalman method’s robustness in handling uncertainties, its ability to work with non-Gaussian and nonlinear dynamics, and its sample-based nature make it a powerful tool for training autoencoders. This integration provides several practical advantages, including improved generalization, resilience to noise, and enhanced performance in capturing the behavior of complex turbulent flows.

The proposed framework will be rigorously tested on turbomachinery separated flows to estimate quantities such as separation location, drag coefficient, lift coefficient, and pressure coefficient. This work’s research objectives include developing the ensemble-Kalman-autoencoder framework, assessing its performance in accurately capturing flow behaviors and its impact on uncertainty quantification and convergence speed, and enhancing confidence in decision-making by improving the understanding of uncertainty quantification.

Through this innovative approach, the proposed work aims to significantly advance the field of turbulence modeling, offering a robust method to quantify and reduce epistemic uncertainty. This, in turn, will inform management decisions and enhance design effectiveness, ensuring that systems perform reliably under diverse and unpredictable conditions.


\bibliography{references.bib}


\renewcommand\theequation{\Alph{section}\arabic{equation}} 
\counterwithin*{equation}{section} 
\renewcommand\thefigure{\Alph{section}\arabic{figure}} 
\counterwithin*{figure}{section} 
\renewcommand\thetable{\Alph{section}\arabic{table}} 
\counterwithin*{table}{section} 

\end{document}